\documentclass[preprint,prd,tightenlines,floatfix,
nofootinbib,eqsecnum,superscriptaddress]{revtex4-1}

\usepackage[T1]{fontenc}		

\usepackage{soul}
\usepackage{xspace,bm,bbm,dsfont}

\usepackage{amsmath,amsfonts,amssymb,amstext,mathrsfs}
\usepackage{amsthm}
\usepackage{mathpazo}

\usepackage[dvips]{graphicx}
\usepackage{epsf,float}
\usepackage{revsymb}

\usepackage{dcolumn}
\usepackage{braket}
\usepackage{color,xcolor}
\usepackage{graphicx}
\usepackage{subfigure}
\usepackage{multirow}
\usepackage{tabularx}
\usepackage{pstricks}
\usepackage[section]{placeins}
\usepackage{booktabs}
\usepackage{array}

\usepackage{hyperref}

\usepackage{slashed}
\usepackage{multirow}

\newcommand{\bqa}{\mbox{\boldmath $q$}_{1}}
\newcommand{\bqb}{\mbox{\boldmath $q$}_{2}}

\newcommand{\bp}{\mbox{\boldmath $p$}}

\usepackage[normalem]{ulem}

\begin{document}


\author{Piotr Lebiedowicz}
\email{Piotr.Lebiedowicz@ifj.edu.pl}
\affiliation{Institute of Nuclear Physics Polish Academy of Sciences, 
Radzikowskiego 152, PL-31342 Krak\'ow, Poland}

\author{Antoni Szczurek
\footnote{Also at \textit{College of Natural Sciences, 
Institute of Physics, University of Rzesz\'ow, 
Pigonia 1, PL-35310 Rzesz\'ow, Poland}.}}
\email{Antoni.Szczurek@ifj.edu.pl}
\affiliation{Institute of Nuclear Physics Polish Academy of Sciences, 
Radzikowskiego 152, PL-31342 Krak\'ow, Poland}

\title{\boldmath Production of $f_2(1270)$ meson in $pp$ collisions at the LHC \\
via gluon-gluon fusion in the $k_t$-factorization approach}

\begin{abstract}
We calculate inclusive cross section for $f_2(1270)$ tensor meson
production via color singlet gluon-gluon fusion
in the $k_t$-factorization approach with unintegrated gluon
distribution functions (UGDFs).
The process may be potentially interesting in the context of searches 
for saturation effects.
The energy-momentum tensor, 
equivalent to helicity-2 coupling, and helicity-0 coupling 
are used for the $g^* g^* \to f_2(1270)$ vertex.
Two somewhat different parametrizations of helicity-2 and helicity-0
tensorial structure from the literature are used in our calculations. 
Some parameters are extracted from 
$\gamma \gamma \to f_2(1270) \to \pi \pi$ reactions.
Different modern UGDFs from the literature are used.
The results strongly depend on the parametrization of the 
$g^* g^* \to f_2(1270)$ form factor. 
Our results for transverse momentum distributions of $f_2$ are compared 
to preliminary ALICE data.
We can obtain agreement with the data only at larger $f_2(1270)$
transverse momenta only for some parametrizations of 
the $g^* g^* \to f_2(1270)$ form factor.
No obvious sign of the onset of saturation is possible.
At low transverse momenta one needs to include 
also the $\pi\pi$ final-state rescattering. 
The agreement with the ALICE data can be obtained
by adjusting probability of formation and survival of $f_2(1270)$ 
in a harsh quark-gluon and multipion environment.
The pomeron-pomeron fusion mechanism is discussed in addition
and results are quantified.
\end{abstract}

\maketitle

\section{Introduction}

The mechanism of $f_2(1270)$ meson production in proton-proton
collisions at high energies was not carefully studied so far.
In the {\tt PYTHIA} event generator $f_2(1270)$ is 
not produced in a primary
fragmentation process but occurs only in decays, e.g. 
$J/\psi \to f_2(1270) \omega$, 
$D_s^{\pm} \to f_2(1270) \pi^{\pm}$,
$B^{\pm} \to \tau^{\pm} \nu_{\tau}/\bar{\nu}_{\tau} f_2(1270)$.
The corresponding branching fractions are rather small so one cannot
expect large contributions.
On the other hand it is rather difficult to observe 
$f_2(1270)$ experimentally. 
The dominant, and relatively easy, decay channel
is $f_2(1270) \to \pi^+ \pi^-$.
Then the signal is on a huge non-reduceable $\pi^+ \pi^-$ background. 
So far only STAR \cite{Adams:2003cc} and ALICE \cite{Lee:thesis} undertook
experimental efforts.
Some time ago \cite{FillionGourdeau:2007ee} it was suggested that the gluon-gluon
fusion could be the dominant production mechanism.
Certainly this interesting working hypothesis requires further
elaboration and experimental confirmation.

In the present paper we follow the idea from 
\cite{FillionGourdeau:2007ee} and try
to shed new light on the situation.
We will apply the $k_t$-factorization approach
successfully used for $\chi_c$ quarkonium production
\cite{Cisek:2017gno,Babiarz:2020jkh},
for $\eta_c(1S,2S)$ production 
\cite{Baranov:2019joi,Babiarz:2019mag},
and recently for $f_{0}(980)$ production \cite{Lebiedowicz:2020bwo}
in proton-proton collisions.
In this study, we focus on the production of $f_{2}(1270)$ meson
in $p p$ collisions.
Recently the production of $f_2$ was also studied
in $\gamma p \to f_2 p$ reaction \cite{Mathieu:2020zpm}.
The $g^* g^* \to f_2(1270)$ vertex is not known a priori.
We will try to verify the hypothesis
of the dominance of the helicity-2 component,
the coupling of spin-2 meson to the energy-momentum tensor
\cite{FillionGourdeau:2007ee}, 
by comparing our results to the preliminary ALICE data \cite{Lee:thesis}.
We shall use modern unintegrated gluon distributions from the literature. 

The tensor-meson dominance for energy-momentum tensor 
(see e.g.~\cite{FillionGourdeau:2007ee}) 
is a possibility used already in 
$\gamma \gamma \to f_2(1270)$ subprocess \cite{Suzuki:1993zs}
for two on-shell photons.
In \cite{Ewerz:2013kda} two tensor structures corresponding to 
$\Gamma^{(0)}$ helicity-0 and $\Gamma^{(2)}$ helicity-2
couplings were found and their strength 
was determined from the comparison to the Belle data
for the $\gamma \gamma \to \pi \pi$ reactions.
The data with the on-shell photons require dominance of 
helicity-2 coupling over helicity-0 coupling; 
see Refs.~\cite{Uehara:2008ep,Pennington:2008xd,Dai:2014zta}.
The $\gamma^* \gamma \to f_2(1270)$ 
coupling was discussed in 
\cite{Achasov:1985ad,Achasov:2015pha,Braun:2000cs,Braun:2016tsk}. 
The $f_2$ meson transition form factors were discussed 
e.g. in \cite{Schuler:1997yw,Hoferichter:2020lap} in the quark model
and in the asymptotic regime of one large virtuality, respectively.
The differential cross section for
the process $\gamma^{*} \gamma \to \pi^{0} \pi^{0}$
in $e^{+}e^{-}$ scattering up to $Q^{2} = 30$~GeV$^{2}$
was studied by the Belle Collaboration 
\cite{Masuda:2015yoh}.
The transition form factor of the $f_{0}(980)$ meson 
and helicity-0, -1, and -2 transition form factors 
of the $f_{2}(1270)$ meson were extracted there.
We will also use tensorial structures 
for the $\gamma^{*} \gamma^{*} \to f_2(1270)$ vertex
from \cite{Pascalutsa:2012pr} 
(see also Ref.~\cite{Poppe:1986dq}). 
Recently, some numerical results 
for the helicity amplitudes 
of $\gamma^{*}\gamma^{*} \to \pi \pi$ that, include $f_{2}(1270)$,
depending of the photon virtualities, 
were presented in \cite{Hoferichter:2019nlq,Danilkin:2019opj}.

\section{Some details of the model calculations}
\label{sec:model}

\subsection{$\gamma^{*} \gamma^{*} \to f_2(1270)$ vertex}
\label{sec:gamgamf2_vertex}

\subsubsection{Ewerz-Maniatis-Nachtmann vertex (EMN)}
\label{sec:EMN}

In \cite{Ewerz:2013kda} the $f_{2} \gamma \gamma$ vertex 
for `on-shell' $f_{2}$ meson
and real photons was considered;
see Eq.~(3.39) of \cite{Ewerz:2013kda} 
and the discussion in Sec.~5.3 therein.
In this approach the photon-photon-$f_2$ vertices come
from Lagrangian formulation for both on-shell photons
and fulfil gauge invariance.
The same model is used then off-mass shell for virtual photons.
   
Here we are interested in $\gamma^*(Q_1^2) \gamma^*(Q_2^2) \to f_2(1270)$ 
process, thus to describe the dependence on photon virtualities
we should introduce the vertex form factors 
$F^{(0)}(Q_1^2,Q_2^2)$ and $F^{(2)}(Q_1^2,Q_2^2)$
for the helicity-0 coupling and the helicity-2 coupling, respectively.

Then the $\gamma^* \gamma^* \to f_2(1270)$ vertex,
including the form factors $F^{(\Lambda)}(Q_1^2,Q_2^2)$,
can be parametrized as
\begin{eqnarray}
\Gamma_{\mu\nu\kappa\lambda}^{(f_2 \gamma \gamma)}(q_1,q_2) 
&=& 2 a_{f_2 \gamma\gamma}\, 
\Gamma_{\mu\nu\kappa\lambda}^{(0)}(q_1,q_2)\,
F^{(0)}(Q_1^2,Q_2^2)  \nonumber \\
&&-b_{f_2 \gamma\gamma}\, 
\Gamma_{\mu\nu\kappa\lambda}^{(2)}(q_1,q_2) \,
F^{(2)}(Q_1^2,Q_2^2)  \,,
\label{EMN_vertex}
\end{eqnarray}
with two rank-four tensor functions,
\begin{eqnarray}
&&\Gamma_{\mu\nu\kappa\lambda}^{(0)} (q_1,q_2) =
\Big[(q_1 \cdot q_2) g_{\mu\nu} - q_{2\mu} q_{1\nu}\Big] 
\Big[q_{1\kappa}q_{2\lambda} + q_{2\kappa}q_{1\lambda} - 
\frac{1}{2} (q_1 \cdot q_2) g_{\kappa\lambda}\Big] \,,
\label{Gam0}\\
&&\Gamma_{\mu\nu\kappa\lambda}^{(2)} (q_1,q_2) = \,
 (q_1\cdot q_2) (g_{\mu\kappa} g_{\nu\lambda} + g_{\mu\lambda} g_{\nu\kappa} )
+ g_{\mu\nu} (q_{1\kappa} q_{2\lambda} + q_{2\kappa} q_{1\lambda} ) \nonumber \\
&& \qquad \qquad \qquad \quad - q_{1\nu} q_{2 \lambda} g_{\mu\kappa} - q_{1\nu} q_{2 \kappa} g_{\mu\lambda} 
- q_{2\mu} q_{1 \lambda} g_{\nu\kappa} - q_{2\mu} q_{1 \kappa} g_{\nu\lambda} 
\nonumber \\
&& \qquad \qquad \qquad \quad - [(q_1 \cdot q_2) g_{\mu\nu} - q_{2\mu} q_{1\nu} ] \,g_{\kappa\lambda} \,;
\label{Gam2}
\end{eqnarray}
see Eqs.~(3.18)--(3.22) of \cite{Ewerz:2013kda}.

To obtain $a_{f_{2} \gamma \gamma}$ 
and $b_{f_{2} \gamma \gamma}$ in (\ref{EMN_vertex})
we use the experimental value
of the radiative decay width 
\begin{eqnarray}
&&\Gamma(f_{2} \to \gamma \gamma) = (2.93 \pm 0.40) \,\mathrm{keV}\,,\nonumber \\
&&\mathrm{helicity\; zero \;contribution} \approx 9 \% \;\mathrm{of}\;\Gamma(f_{2} \to \gamma \gamma)\,,
\label{PDG_values}
\end{eqnarray}
as quoted for the preferred solution III 
in Table~3 of \cite{Dai:2014zta}.
Using the decay rate from (5.28) of \cite{Ewerz:2013kda}
\begin{equation}
\Gamma(f_{2} \to \gamma \gamma) =
\frac{m_{f_{2}}}{80 \pi} 
\left( \frac{1}{6} m_{f_{2}}^{6} |a_{f_{2} \gamma \gamma}|^{2} + m_{f_{2}}^{2} |b_{f_{2} \gamma \gamma}|^{2} \right)\,,
\label{decay_rate}
\end{equation}
and assuming $a_{f_{2} \gamma \gamma} > 0$
and $b_{f_{2} \gamma \gamma} >0$, we find
\begin{eqnarray}
&&a_{f_{2} \gamma \gamma}=
\alpha_{\rm em}\, \times\,
1.17\;\mathrm{GeV^{-3}}\,,
\label{a_coupling}
\\
&&b_{f_{2} \gamma \gamma}=
\alpha_{\rm em}\, \times\,
2.46\;\mathrm{GeV^{-1}}\,,
\label{b_coupling}
\end{eqnarray}
where $\alpha_{\rm em} = e^{2}/(4 \pi) \simeq 1/137$
is the electromagnetic coupling constant.

\subsubsection{Pascalutsa-Pauk-Vanderhaeghen vertex (PPV)}
\label{sec:PPV}

In Refs.~\cite{Poppe:1986dq,Schuler:1997yw,Pascalutsa:2012pr} 
it was shown that the most general amplitude for
the process
$\gamma^\ast (q_1, \lambda_1) + \gamma^\ast(q_2, \lambda_2) \to f_{2}(\Lambda)$, 
describing the transition from an initial state of two  
virtual photons to a tensor meson $f_{2}$ ($J^{PC} = 2^{++}$)
with the mass $m_{f_{2}}$ and 
helicity $\Lambda = \pm 2, \pm 1, 0$, 
involves five independent structures 
(invariant amplitudes).

In the formalism presented in \cite{Pascalutsa:2012pr} 
the $\gamma^* \gamma^* \to f_2(1270)$ vertex 
was parametrized as
\begin{eqnarray}
&&\Gamma_{\mu\nu\kappa\lambda}^{(f_2 \gamma \gamma)}(q_1,q_2) 
= 4 \pi \alpha_{\rm em}
\left\{
 \left[ R_{\mu \kappa} (q_1, q_2) R_{\nu \lambda} (q_1, q_2) 
+ \frac{s}{8 X} \, R_{\mu \nu}(q_1, q_2) 
(q_1 - q_2)_\kappa \, (q_1 - q_2)_\lambda \right]
\right. 
\nonumber \\
&&\quad \quad \left. 
\times \frac{\nu}{m_{f_{2}}} T^{(2)}(Q_1^2, Q_2^2)
\right. 
\nonumber \\
&&\quad \quad \left. + R_{\nu \kappa}(q_1, q_2) (q_1 - q_2)_\lambda  
\left( q_{1 \mu} + \frac{Q_1^2}{\nu} q_{2 \mu} \right) 
\frac{1}{m_{f_{2}}} 
T^{(1)}(Q_1^2, Q_2^2) 
\right. 
\nonumber \\
&&\quad \quad \left. + R_{\mu \kappa}(q_1, q_2) (q_2 - q_1)_\lambda   
\left( q_{2 \nu} + \frac{Q_2^2}{\nu} q_{1 \nu} \right) 
\frac{1}{m_{f_{2}}} 
T^{(1)}(Q_2^2, Q_1^2)
\right. 
\nonumber \\ 
&&\quad \quad \left. + R_{\mu \nu}(q_1, q_2) (q_1 - q_2)_\kappa \, (q_1 - q_2)_\lambda \, \frac{1}{m_{f_{2}}} 
T^{(0, {\rm T})}(Q_1^2, Q_2^2) 
\right. 
\nonumber \\
&&\quad \quad \left. + \left( q_{1 \mu} + \frac{Q_1^2}{\nu} q_{2 \mu} \right) 
\left( q_{2 \nu} + \frac{Q_2^2}{\nu} q_{1 \nu} \right)  
(q_1 - q_2)_\kappa  (q_1 - q_2)_\lambda 
\frac{1}{m_{f_{2}}^3}
T^{(0, {\rm L})}(Q_1^2, Q_2^2)
\right\},
\label{PPV_vertex}
\end{eqnarray}
where photons with four-momenta $q_{1}$ and $q_{2}$ have
virtualities,
$Q_{1}^{2}=-q_{1}^{2}$ and $Q_{2}^{2}=-q_{2}^{2}$,
$s = (q_{1}+q_{2})^{2} = 2 \nu - Q_1^2 - Q_2^2$, 
$X = \nu^{2} - q_1^2 q_2^2$,
$\nu = (q_{1} \cdot q_{2})$, and
\begin{equation}
R_{\mu \nu}(q_{1},q_{2}) = -g_{\mu \nu} + \frac{1}{X}
\left[
\nu \left( q_{1 \mu} q_{2 \nu} + q_{2 \mu} q_{1 \nu} \right)
- q_1^2 q_{2 \mu} q_{2 \nu} - q_2^2 q_{1 \mu} q_{1 \nu}
\right]\,.
\end{equation}

In Eq.~(\ref{PPV_vertex}) $T^{(\Lambda)}(Q_1^2, Q_2^2)$ 
are the $\gamma^* \gamma^* \to f_2(1270)$ 
transition form factors for $f_2(1270)$ helicity $\Lambda$. 
For the case of helicity zero, 
there are two form factors depending on whether both photons are transverse (superscript ${\rm T}$) 
or longitudinal (superscript ${\rm L}$).

We can express the transition form factors as
\begin{eqnarray}
T^{(\Lambda)}(Q_1^2, Q_2^2) = F^{(\Lambda)}(Q_1^2, Q_2^2) \, T^{(\Lambda)}(0,0) \,.
\label{TFF_aux}
\end{eqnarray}

In the limit $Q_{1,2}^{2} \to 0$ 
only $T^{(0,{\rm T})}$ and $T^{(2)}$ contribute
and their values at $Q_{1,2}^{2} \to 0$ determine
the two-photon decay width of $f_{2}(1270)$ meson.

Comparing the two approaches given by 
(\ref{EMN_vertex}) and (\ref{PPV_vertex})--(\ref{TFF_aux})
for both real photons ($Q_{1}^{2} = Q_{2}^{2} = 0$) 
and at $\sqrt{s} = m_{f_{2}}$
we found the correspondence 
\begin{eqnarray}
&&4 \pi \alpha_{\rm em} \, T^{(0,{\rm T})}(0,0) =  
- a_{f_{2} \gamma \gamma} \, \frac{m_{f_{2}}^{3}}{2}\,,\\
&&4 \pi \alpha_{\rm em} \, T^{(2)}(0,0) = 
- b_{f_{2} \gamma \gamma} \, 2 m_{f_{2}} \,.
\label{PPV_parameters}
\end{eqnarray}
%

\subsection{$g^* g^* \to f_2(1270)$ vertex}
\label{sec:ggf2_vertex}

We will apply the formalism 
for the $\gamma^* \gamma^* \to f_2$ vertices
discussed in Sec.~\ref{sec:gamgamf2_vertex}.
This means that the $g^* g^* \to f_2(1270)$ vertex
has the same form as that for the $\gamma^* \gamma^* \to f_2(1270)$ vertex,
but with the replacement (\ref{replacement}).

Because $f_2(1270)$ is extended, finite size object one can expect 
in addition a form factor(s) $F(Q_{1}^{2}, Q_{2}^{2})$ 
associated with the gluon virtualities for
the $g^* g^* \to f_2$ vertex\footnote{In general 
the form factors for different tensorial structures can be different.}.
In the present letter the form factor, 
identical for $\Lambda = 0$ and $\Lambda = 2$, 
is parametrized in different ways as:
\begin{eqnarray}
F(Q_1^2,Q_2^2)&=& 
\frac{\Lambda_M^2}{Q_1^2+Q_2^2+\Lambda_M^2} \,,
\label{monopole} \\
F(Q_1^2,Q_2^2)&=& 
\left( \frac{\Lambda_D^2}{Q_1^2+Q_2^2+\Lambda_D^2} \right)^2 \,,
\label{dipole} \\ 
F(Q_1^2,Q_2^2)&=&   
\frac{\Lambda_{1}^2}{Q_1^2+\Lambda_{1}^2}
\frac{\Lambda_{1}^2}{Q_2^2+\Lambda_{1}^2}\,,
\label{ffa} \\
F(Q_1^2,Q_2^2)&=& 
\frac{\Lambda_{2}^4}{(Q_1^2+\Lambda_{2}^2)^{2}}
\frac{\Lambda_{2}^4}{(Q_2^2+\Lambda_{2}^2)^{2}}\,,
\label{ffb}                   
\end{eqnarray}
where $\Lambda$'s above are parameters whose value 
is expected to be close to the resonance mass \cite{Schuler:1997yw}.
In the case of non-factorized form factors, 
monopole (\ref{monopole}) and dipole (\ref{dipole}), 
we use $\Lambda_M = \Lambda_D = m_{f_{2}}$.
The values of $\Lambda_1$ in (\ref{ffa}) 
and $\Lambda_2$ in (\ref{ffb})
are expected to be of order of 1~GeV.
The results for different forms of the form factors (\ref{monopole})--(\ref{ffb})
are presented in Fig.~\ref{fig:dN_dpt_ff}.
The results strongly depend on the parametrization chosen
and the value of the corresponding parameter.

\subsection{$k_t$-factorization approach}

\begin{figure}[!ht]
\begin{center}
\includegraphics[width=0.35\textwidth]{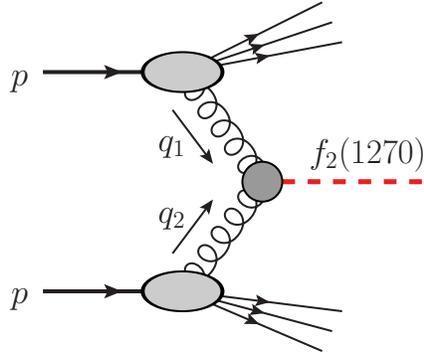}
\caption{\label{fig:diagram_gg_f2}
\small
General diagram for inclusive $f_2(1270)$ production
via gluon-gluon fusion in proton-proton collisions.}
\end{center}
\end{figure}

In Fig.~\ref{fig:diagram_gg_f2} we show a generic Feynman diagram for 
$f_2(1270)$ meson production in proton-proton collision via 
gluon-gluon fusion.
This diagram illustrates the situation adequate for 
the $k_t$-factorization calculations used in the present paper.

The differential cross section for inclusive $f_{2}(1270)$ meson production 
via the $g^* g^* \to f_{2}(1270)$ fusion in the $k_t$-factorization approach
can be written as:
\begin{eqnarray}
\frac{d\sigma}{dy d^2\bp} = \int {d^2 \bqa \over \pi \bqa^2} 
{\cal F}_{g}(x_1,\bqa^2) \int {d^2 \bqb \over \pi \bqb^2}  
{\cal F}_{g}(x_2,\bqb^2) \, \delta^{(2)} (\bqa + \bqb - \bp ) \, 
{\pi \over (x_1 x_2 s)^2} \overline{|{\cal M}_{g^{*} g^{*} \to f_2}|^2} \,.\qquad \quad
\label{eq:cross_section_2}
\end{eqnarray}
Here $\bqa$, $\bqb$ and $\bp$ denote the transverse momenta of the
gluons and the $f_{2}(1270)$ meson, respectively.
The $f_{2}$ meson is on-shell and its momentum satisfies
$p^{2} = m_{f_{2}}^{2}$.
${\cal M}_{g^{*} g^{*} \to f_2}$ is the matrix element 
for off-shell gluons for the hard subprocess and ${\cal F}_{g}$ are the gluon unintegrated 
distribution functions (UGDFs) for colliding protons. 
The UGDFs depend on gluon longitudinal momentum fractions
$x_{1,2} = m_{T} \exp(\pm {\rm y})/\sqrt{s}$
and $\bqa^2, \bqb^2$ entering the hard process.
In principle, they can depend also on factorization scales 
$\mu_{F,i}^2$, $i = 1, 2$.
It is reasonable to assume $\mu_{F,1}^2 = \mu_{F,2}^2 = m_{T}^2$.
Here $m_{T}$ is transverse mass of the produced $f_{2}(1270)$ meson;
$m_{T} = \sqrt{\bp^{2} + m_{f_{2}}^{2}}$.
The~$\delta^{(2)}$ function in Eq.~(\ref{eq:cross_section_2}) 
can be easily eliminated by introducing
$\bqa + \bqb$ and $\bqa - \bqb$ transverse momenta \cite{Cisek:2017gno}.

The off-shell matrix element can be written as (we restore the
color-indices $a$ and $b$)
\begin{eqnarray}
{\cal{M}}^{ab} = {q_{1t}^\mu q_{2t}^\nu \over |\bqa| |\bqb|}
{\cal{M}}^{ab}_{\mu \nu}  = 
{q_{1+} q_{2-} \over |\bqa| |\bqb|} n^{+\mu} n^{-\nu}
{\cal{M}}^{ab}_{\mu \nu} = 
{x_1 x_2 s \over 2 |\bqa| |\bqb| } n^{+\mu} n^{-\nu} {\cal{M}}^{ab}_{\mu \nu}
\end{eqnarray}
with the lightcone components of gluon momenta
$q_{1+} = x_{1} \sqrt{s/2}$, $q_{2-} = x_{2} \sqrt{s/2}$.
Here the matrix-element reads
\begin{eqnarray}
{\cal{M}}_{\mu \nu} = 
\Gamma_{\mu\nu\kappa\lambda}^{(f_2 gg)}(q_{1t}, q_{2t}) \, 
(\epsilon^{(f_{2})\,\kappa \lambda}(p))^{*}\,,
\end{eqnarray}
where $\epsilon^{(f_{2})}$ is the polarization tensor for the $f_{2}(1270)$ meson.

In the $k_t$-factorization approach in \cite{FillionGourdeau:2007ee} the matrix element 
squared (for energy-momentum tensor coupling) was written as:
\begin{eqnarray}
&&\overline{ | {\cal M}_{g^{*} g^{*} \to f_2} |^2}
\nonumber \\
&&\quad = 
\frac{1}{(N_{c}^{2}-1)^{2}} \sum_{a,b} 
\frac{q_{1t\,\mu_1}}{q_{1t}} \frac{q_{2t\,\nu_1}}{q_{2t}}
V_{ab}^{\alpha_1 \beta_1 \mu_1 \nu_1}(q_{1},q_{2})\,
P^{(2)}_{\alpha_1 \beta_1, \alpha_2 \beta_2}(p)
\frac{q_{1t\,\mu_2}}{q_{1t}} \frac{q_{2t\,\nu_2}}{q_{2t}}
\left(V_{ab}^{\alpha_2 \beta_2 \mu_2 \nu_2}(q_{1},q_{2}) \right)^{*}
\nonumber \\
&&\quad = 
\frac{1}{(N_{c}^{2}-1) \kappa^{2}}
P^{(2)}_{\alpha_1 \beta_1, \alpha_2 \beta_2}(p)
H_{\perp}^{\alpha_1 \beta_1}(q_{1t}, q_{2t}) 
H_{\perp}^{\alpha_2 \beta_2}(q_{1t}, q_{2t}) \, 
\left( \frac{x_{1}x_{2}s}{2q_{1t}q_{2t}} \right)^{2}\,,
\label{general_M2}
\end{eqnarray}
where $N_{c}$ is the number of colors,
$V_{ab}^{\alpha \beta \mu \nu}$ 
is the $g g \to f_{2}$ vertex\footnote{
Please note that
the order of Lorentz indices here 
(and in Ref.~\cite{FillionGourdeau:2007ee})
is different than in Eq.~(\ref{Gam2}).} 
(see Eq.~(A1) of \cite{FillionGourdeau:2007ee}),
and $\kappa \approx {\cal O}(0.1\, {\rm GeV})$ 
is to be fixed by experiment.
The explicit forms for the spin-2 projector $P^{(2)}$
and $H^{\alpha \beta}_{\perp}$ functions (with transverse
components) are given in \cite{FillionGourdeau:2007ee}.
In the above formula (\ref{general_M2}) 
$\alpha_{\rm s}$ is not explicit 
but is hidden in the normalization constant.
In our calculation we will make $\alpha_{\rm s}$ explicit, i.e. include its running
with relevant scales.
We have checked that the approach in \cite{FillionGourdeau:2007ee}
is equivalent to the approach
with the helicity-2 EMN vertex function (\ref{Gam2})
when ignoring running of $\alpha_{\rm s}$
and vertex form factor $F(Q_1^2,Q_2^2)$.
Having $F(Q_1^2,Q_2^2)$ is crucial for description
of transverse momentum distribution
of $f_{2}(1270)$ as will be discussed in the result section.

The $g^* g^* \to f_{2}(1270)$ coupling entering in the matrix element squared
can be obtained from that for the $\gamma^* \gamma^* \to f_{2}(1270)$ coupling
by the following replacement:
\begin{equation}
\alpha_{\rm{em}}^2 \to \alpha_{\rm s}^2  \,
\frac{1}{4 N_c (N_c^2 - 1)} \,
\frac{1}{(<e_q^2>)^2} \,.
\label{replacement}
\end{equation}
Here $(<e_{q}^2>)^2 = 25/162$ for 
the $\frac{1}{\sqrt{2}} \left(u \bar u + d \bar d \right)$
flavor structure assumed for $f_{2}(1270)$.

In realistic calculations the running of strong coupling constants
must be included.
In our numerical calculations presented below
the renormalization scale is taken 
in the form:
\begin{equation}
\alpha_{\rm s}^2 \to 
\alpha_{\rm s}(\max{\{m_{T}^2,\bqa^2\}})\,
\alpha_{\rm s}(\max{\{m_{T}^2,\bqb^2\}})\,.
\label{alpha_s}
\end{equation}

The Shirkov-Solovtsov prescription \cite{Shirkov:1997wi} is used to extrapolate down to small renormalization scales 
relevant for the $f_2(1270)$ production for the ALICE kinematics.

\subsection{A simple $\pi\pi$ final-state rescattering model}
\begin{figure}[!ht]
\begin{center}
\includegraphics[width=0.45\textwidth]{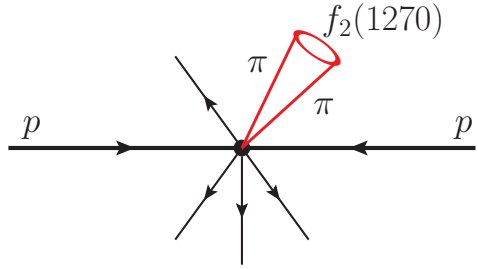}
\caption{\label{fig:rescattering}
\small
General diagram for the $\pi \pi$ final-state rescattering leading to $f_2(1270)$ production
in proton-proton collisions.}
\end{center}
\end{figure}

As will be shown in the present paper the $g^* g^* \to f_2(1270)$
mechanism is insufficient at low $f_2(1270)$ transverse momenta
therefore we consider also a final-state rescattering of produced pions.
The general diagram representing the $\pi \pi$ rescattering is shown in Fig.~\ref{fig:rescattering}.
Both $\pi^+ \pi^-$ and $\pi^0 \pi^0$ rescatterings may lead to
the production of the $f_2(1270)$ meson
as an effect of final state resonance interactions.

The distribution of pions will be not calculated here but instead we will use
a L{\'e}vy parametrization of the inclusive $\pi^0$ cross section
proposed in \cite{Abelev:2012cn} for $\sqrt{s} = 7$~TeV.
At the ALICE energies and midrapidities we assume the following relation:
\begin{equation}
\frac{d \sigma^{\pi^+}}{d{\rm y} d p_t}({\rm y},p_t) =
\frac{d \sigma^{\pi^-}}{d{\rm y} d p_t}({\rm y},p_t) =
\frac{d \sigma^{\pi^0}}{d{\rm y} d p_t}({\rm y},p_t) 
\label{isospin_symmetry}
\end{equation}
to be valid.

Our approach here is similar in spirit to color evaporation approach
considered, e.g., in \cite{Maciula:2018bex,Lebiedowicz:2020bwo}.
In our approach here we do not include possible $\pi \pi$ correlation
functions. They are discussed usually at very small relative momentum.
For identical particles ($\pi^0 \pi^0$ in our case) this is discussed 
usually in the context of Bose-Einstein correlations.
The non-identical particle correlations ($\pi^+ \pi^-$ in our case)
is less popular but also very interesting \cite{Pratt:2003ar,Pratt:2006jf}.
To form the resonance the two pions must be produced in the $\pi \pi$ 
invariant mass window corresponding to the $f_2(1270)$ meson and close 
in space one to each other. Including explicitly the second condition would 
require knowledge of the space-time development of the hadronization 
process and goes far beyond the present study devoted to 
the $g^* g^* \to f_2(1270)$ mechanism. 
Instead we write the number of produced $f_2(1270)$ per event as
\begin{equation}
N^{f_{2}} =
\int d {\rm y}_1 d p_{1t} \int d {\rm y}_2 d p_{2t} \int 
\frac{d \phi_1}{2 \pi} \frac{d \phi_2}{2 \pi}
\frac{dN^{\pi}}{d{\rm y}_1 dp_{1t}} 
\frac{dN^{\pi}}{d{\rm y}_2 dp_{2t}}
P_{\pi \pi \to f_2} \,,
\label{FSR_formula}
\end{equation}
where $dN^{\pi}/(d{\rm y} dp_{t})$ is number of pions
per interval of rapidity and transverse momentum.
Here for $\frac{d N^{\pi}}{d {\rm y}_1 d p_{1t}}$ and
         $\frac{d N^{\pi}}{d {\rm y}_2 d p_{2t}}$          
we use the Tsallis parametrization of $\pi^{0}$ at $\sqrt{s} = 7$~TeV
from \cite{Abelev:2012cn}; 
see~Eq.~(2)~of~\cite{Abelev:2012cn} 
and fit parameters in Table~{3} therein.
In Eq.~(\ref{FSR_formula})
$P_{\pi \pi \to f_2}$ parametrizes probability of the $\pi^+ \pi^-$
and $\pi^0 \pi^0$ formation of $f_2(1270)$ as well as probability of 
its survival in a dense hadronic system. It will be treated here as 
a free parameter adjusted to the $f_2(1270)$ data from \cite{Lee:thesis}.
The distribution $dN^{f_{2}}/(d{\rm y} dp_{t})$ is obtained 
then by calculating ${\rm y}$ and $p_t$ of the $f_2(1270)$ meson 
and binning in these variables. 

The effect of hadronic rescattering 
in high-energy $pp$ collisions was discussed very recently in
\cite{Sjostrand:2020gyg} and the application 
is being developed and will be implemented to {\tt PYTHIA} event generator.

\section{Numerical results}
\label{sec:results}

To convert to the number of $f_2(1270)$ mesons per event, 
as was presented in Ref.~\cite{Lee:thesis}, 
we use the following relation:
\begin{equation}
\frac{d N}{d p_t} = \frac{1}{\sigma_{\rm inel}} \frac{d \sigma}{d p_t} \,.
\label{dN_dpt}
\end{equation}
The inelastic cross section for $\sqrt{s} = 7$~TeV 
was measured at the LHC and is:
\begin{eqnarray}
\sigma_{\rm inel} &=& 73.15 \pm 1.26 \,{\rm(syst.)}\, {\rm mb}
\,,\\
\sigma_{\rm inel} &=& 71.34 \pm 0.36 \,{\rm(stat.)} \pm 0.83 \,{\rm(syst.)} \, {\rm mb} \,,
\label{sigma_ine}
\end{eqnarray}
as obtained by the TOTEM \cite{Antchev:2013gaa} 
and ATLAS \cite{Aad:2014dca} collaborations, respectively. 
In our calculations we take $\sigma_{{\rm inel}} = 72.5$~mb.

\begin{figure}[!ht]
\includegraphics[width=0.495\textwidth]{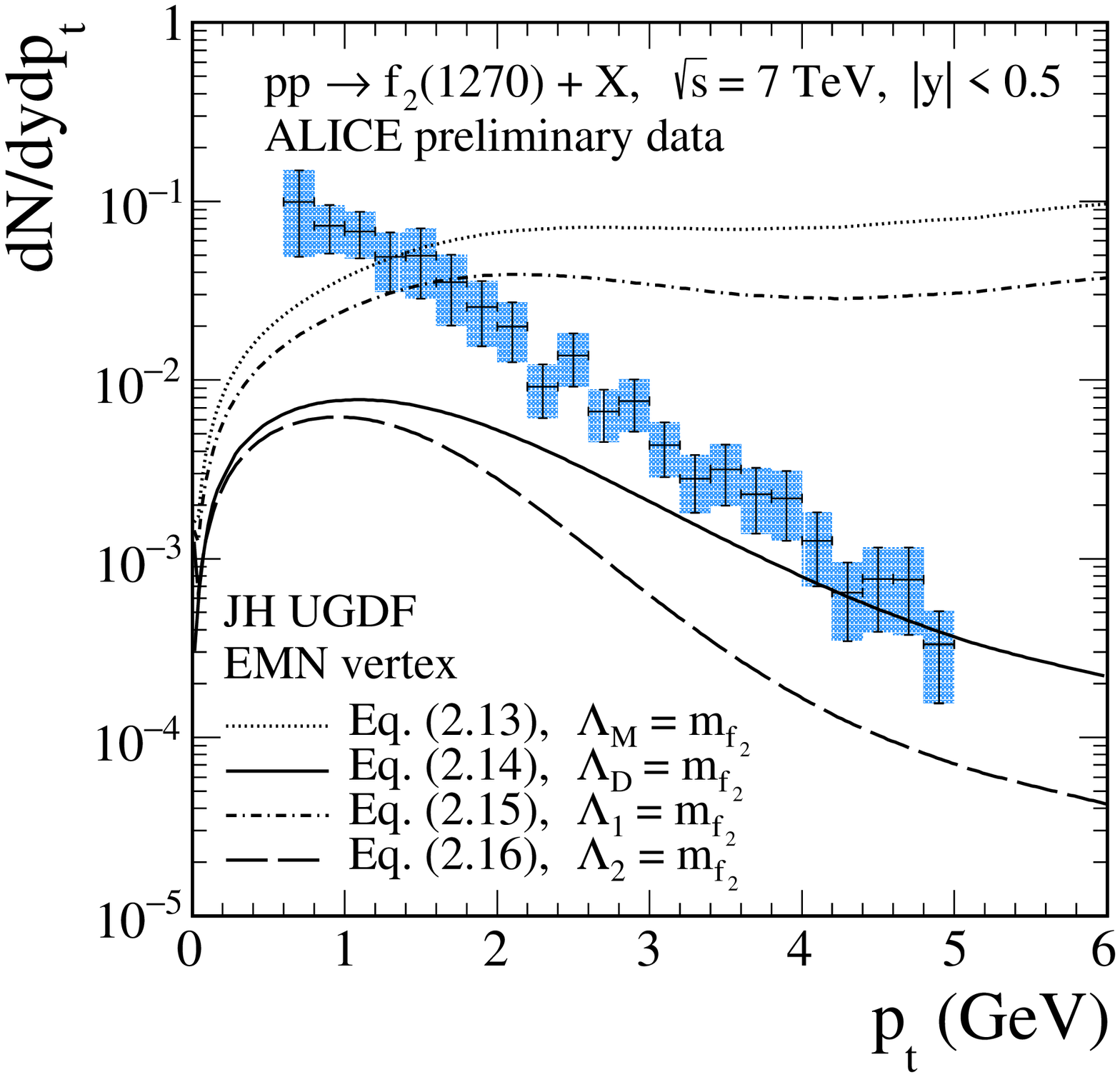}
\includegraphics[width=0.495\textwidth]{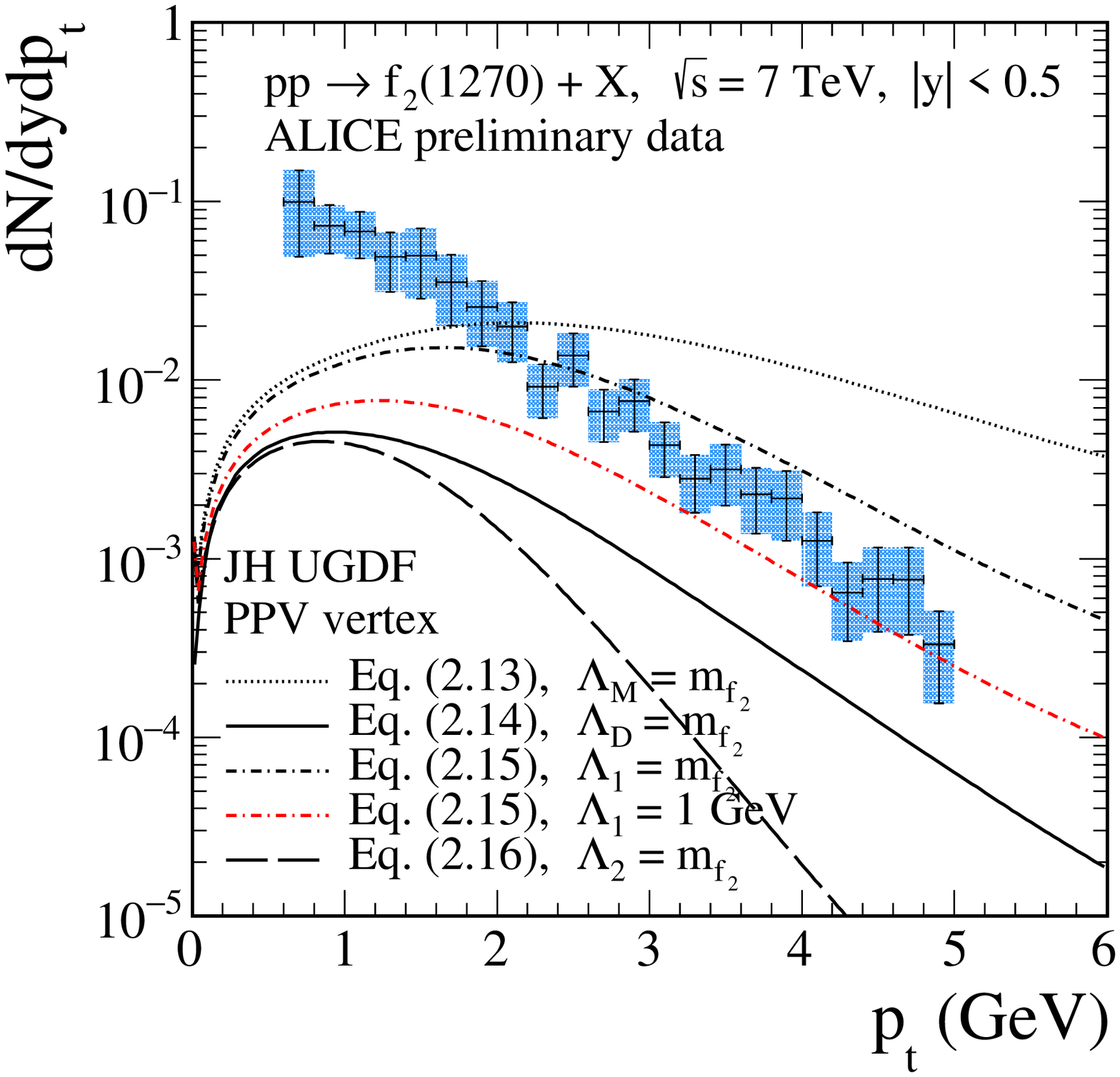}
\caption{\label{fig:dN_dpt_ff}
\small
The $f_{2}(1270)$ meson transverse momentum distributions at
$\sqrt{s}=7$~TeV and $|{\rm y}|<0.5$.
The preliminary ALICE data from \cite{Lee:thesis} 
are shown for comparison.
The results for the EMN (left panel)
and PPV (right panel) $g^* g^* \to f_2(1270)$ vertex
for different parametrizations of $F(Q_1^2,Q_2^2)$ form factor 
(\ref{monopole})--(\ref{ffb}) are shown.
In this calculation the JH UGDF was used.}
\end{figure}

In Fig.~\ref{fig:dN_dpt_ff} we present 
the $f_2(1270)$ meson transverse momentum distributions at
$\sqrt{s}=7$~TeV and $|{\rm y}|<0.5$
together with the preliminary ALICE data from \cite{Lee:thesis}.
Here, for the color-singlet gluon-gluon fusion mechanism,
we used the JH UGDF from \cite{Hautmann:2013tba}.\footnote{This type of UGD has been obtained 
by Hautmann and Jung \cite{Hautmann:2013tba}
from a description of precise HERA data 
on deep inelastic structure function 
by a solution of the CCFM evolution equation
\cite{Ciafaloni:1987ur,Catani:1989yc,Catani:1989sg}.
This UGDF is available
from the {\tt CASCADE} Monte Carlo code 
\cite{Jung:2010si}. 
We use ``JH-2013-set2'' of Ref.~\cite{Hautmann:2013tba},
which we label as ``JH UGDF''.}
We show results for two different $g^{*}g^{*} \to f_{2}$ vertices 
discussed in Sec.~\ref{sec:gamgamf2_vertex}, 
EMN (left panel) and PPV (right panel),
and for different forms of parametrization form factor $F(Q_1^2,Q_2^2)$ 
given by Eqs.~(\ref{monopole})--(\ref{ffb}) and (\ref{TFF_aux}).
The results strongly depend on the parametrization of the form factor.
Assuming the cut-off parameter to be close to the $f_{2}(1270)$ mass 
the forms (\ref{monopole}) and (\ref{ffa}) can be excluded
as they overestimates the ALICE data at larger $p_{t}$.

\begin{figure}[!ht]
\includegraphics[width=0.495\textwidth]{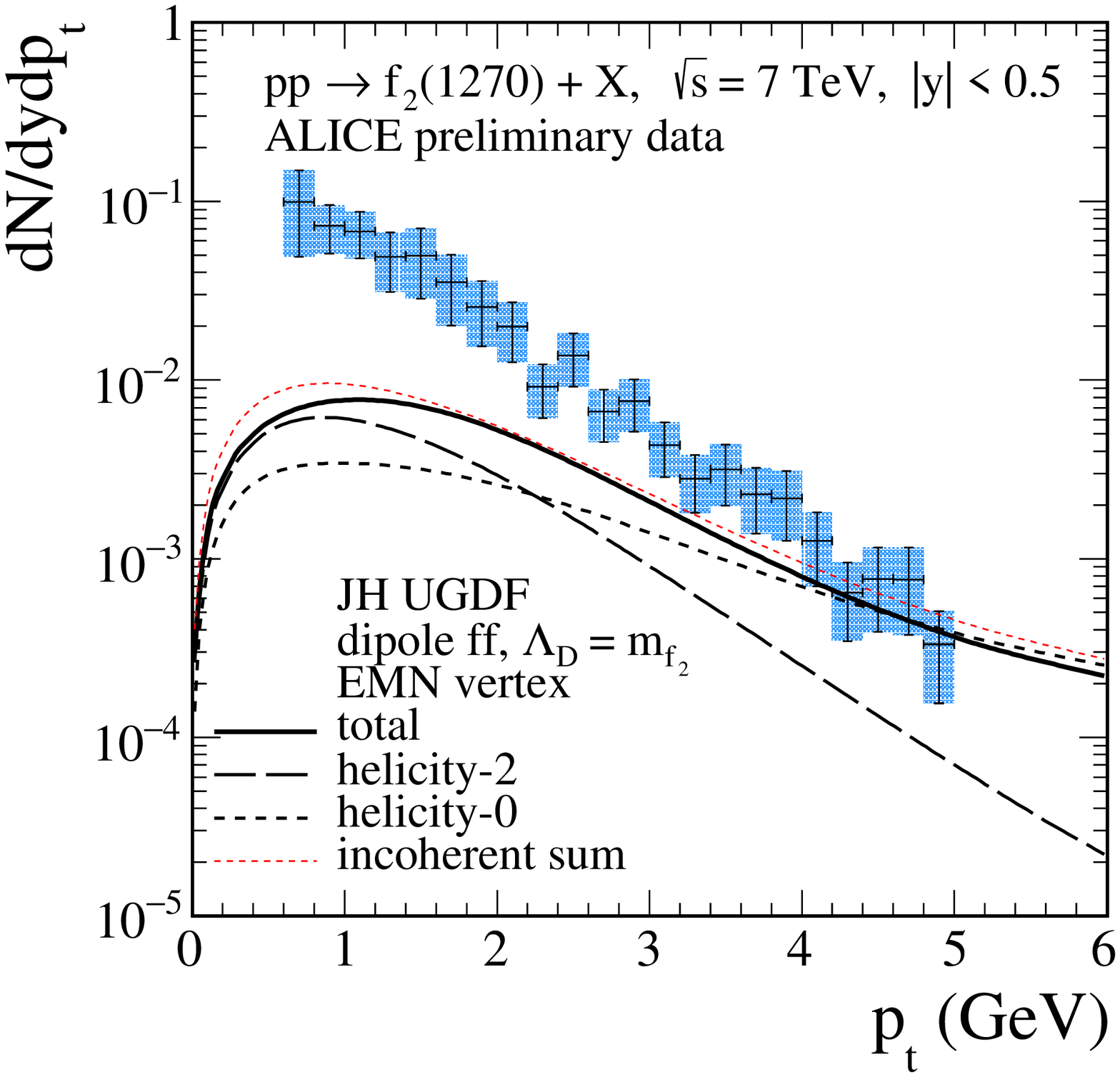}
\includegraphics[width=0.495\textwidth]{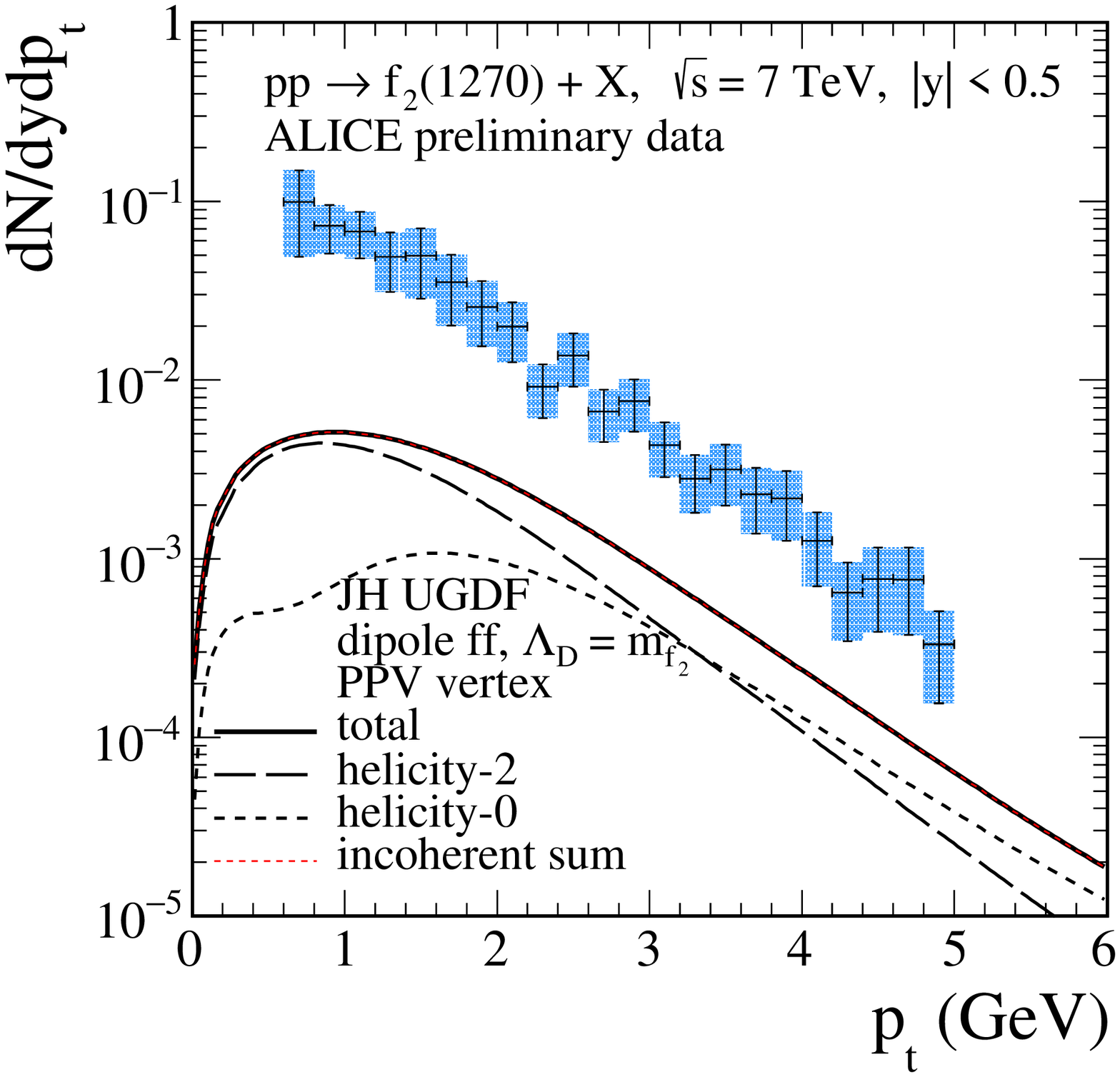}
\caption{\label{fig:dN_dpt_deco}
\small
The $f_{2}(1270)$ meson transverse momentum distributions at
$\sqrt{s}=7$~TeV and $|{\rm y}|<0.5$
together with the preliminary ALICE data from \cite{Lee:thesis}.
Shown are the results calculated in the two approaches, 
EMN (left panel) and PPV (right panel), for
the helicity-0 and helicity-2 components separately
and their coherent sum (total). 
The dotted line corresponds to incoherent sum of the two helicity components.
In this calculation we used dipole form factor parametrization (\ref{dipole})
with $\Lambda_D = m_{f_{2}}$.}
\end{figure}

Fig.~\ref{fig:dN_dpt_deco} shows that
there is some difference in the role of 
$\Lambda = 0, 2$ contributions
for the EMN and PPV vertices.
In the formalism of \cite{Pascalutsa:2012pr} 
[see the PPV vertex (\ref{PPV_vertex})]
there is no interference between 
so-called $\Lambda = 0, {\rm{T}}$ and $\Lambda = 2$ terms while 
the naive use of the formalism from \cite{Ewerz:2013kda} 
[see the EMN vertex (\ref{EMN_vertex})]
generates some interference effects.
Different couplings (independent invariant amplitudes) 
lead to different shapes of the transverse momentum distributions. 
The shape could be verified by experimental data.

\begin{figure}[!ht]
\includegraphics[width=0.495\textwidth]{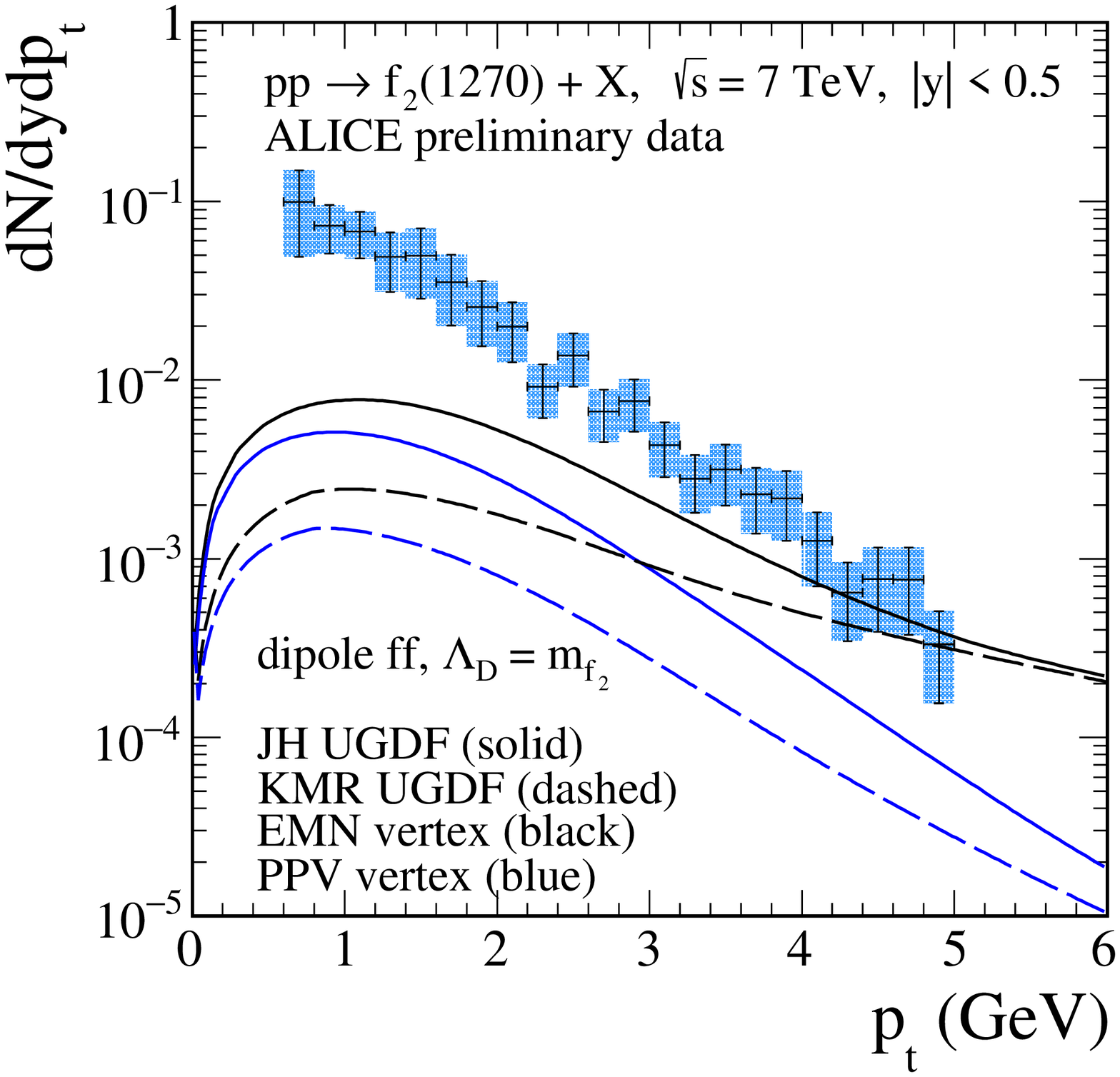}
\includegraphics[width=0.495\textwidth]{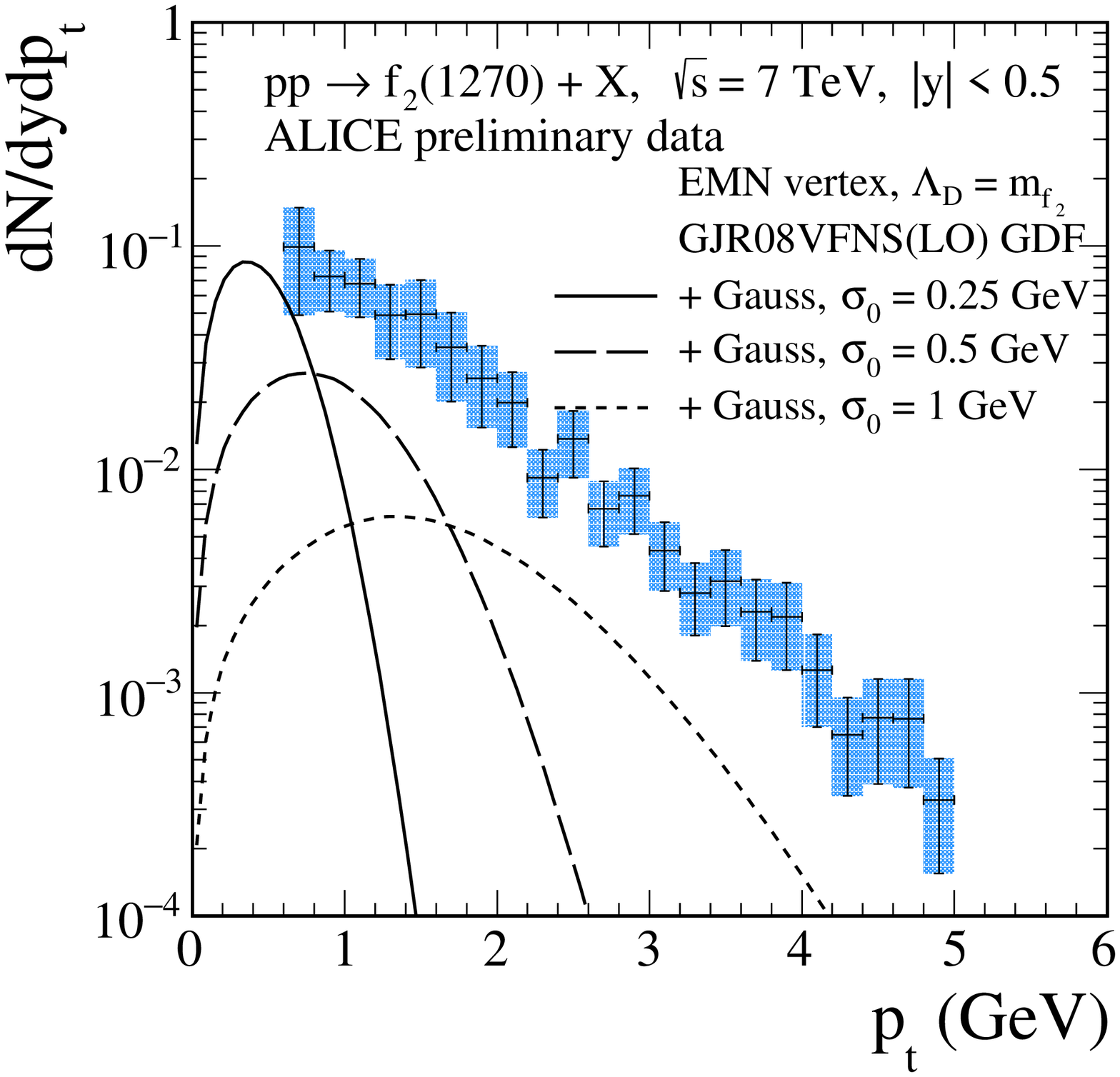}
\caption{\label{fig:GJR}
\small
The $f_{2}(1270)$ meson transverse momentum distributions at
$\sqrt{s}=7$~TeV and $|{\rm y}|<0.5$
together with the preliminary ALICE data from \cite{Lee:thesis}.
In the left panel results for two different UGDFs, 
JH (solid lines) and KMR (dashed lines), are shown. 
In the right panels we show the dependence 
on the Gaussian smearing parameter $\sigma_{0}$
for the GJR08VFNS(LO) GDF \cite{Gluck:2008gs}.
Here the EMN vertex discussed in Sec.~\ref{sec:EMN} 
and the dipole form factor (\ref{dipole}) 
with $\Lambda_D = m_{f_{2}}$ were used.}
\end{figure}

In the left panel of Fig~\ref{fig:GJR} we show results 
for the KMR UGDF.\footnote{Here we use a glue constructed 
according to the prescription
initiated in \cite{Kimber:2001sc} and later updated in 
\cite{Watt:2003vf,Martin:2009ii},
which we label as ``KMR UGDF''.
The KMR UGDF is available from the {\tt CASCADE} Monte Carlo code 
\cite{Jung:2010si}.}
The KMR UGDF (dashed lines) gives smaller cross section 
than the JH UGDF (solid lines).
The results for both UGDFs coincide for large $p_{t}$.
The larger the $f_2(1270)$ transverse momentum
the larger the range of gluon transverse momenta 
$q_{1t}$ and/or $q_{2t}$ are probed.
This means that at larger $f_2$ transverse momenta one enters
a more perturbative region.

In the right panel of Fig~\ref{fig:GJR} 
we show the results with the Gaussian smearing of collinear GDF, 
often used in the context of TMDs, 
for different smearing parameter $\sigma_{0} = 0.25$, 0.5, 1.0~GeV.
The GJR08VFNS(LO) collinear GDF \cite{Gluck:2008gs} 
was used for this purpose.
As expected the shape of $d\sigma/dp_t$ strongly depends on
the value of the smearing parameter $\sigma_{0}$ used in the calculation.
The speed of $d\sigma/dp_t$ approaching to zero for $p_t \to 0$
strongly depends on the value of $\sigma_{0}$. 
It is impossible to describe simultaneously $p_t < 1$~GeV 
and $p_t > 1$~GeV regions with the same value of $\sigma_{0}$. 
This illustrates the generic situation with all UGDFs.

\begin{figure}[!ht]
\begin{center}
\includegraphics[width=0.47\textwidth]{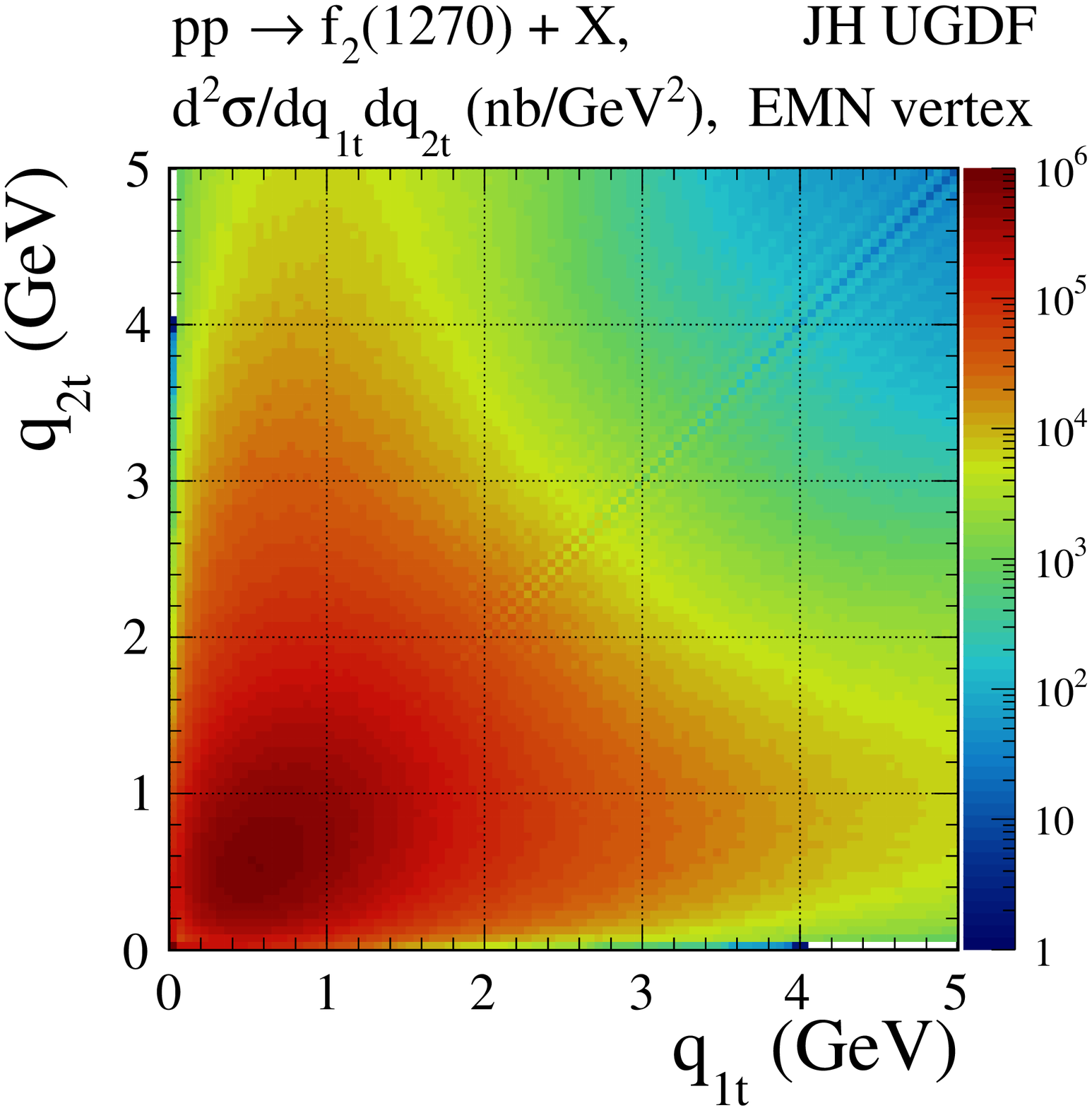}
\includegraphics[width=0.47\textwidth]{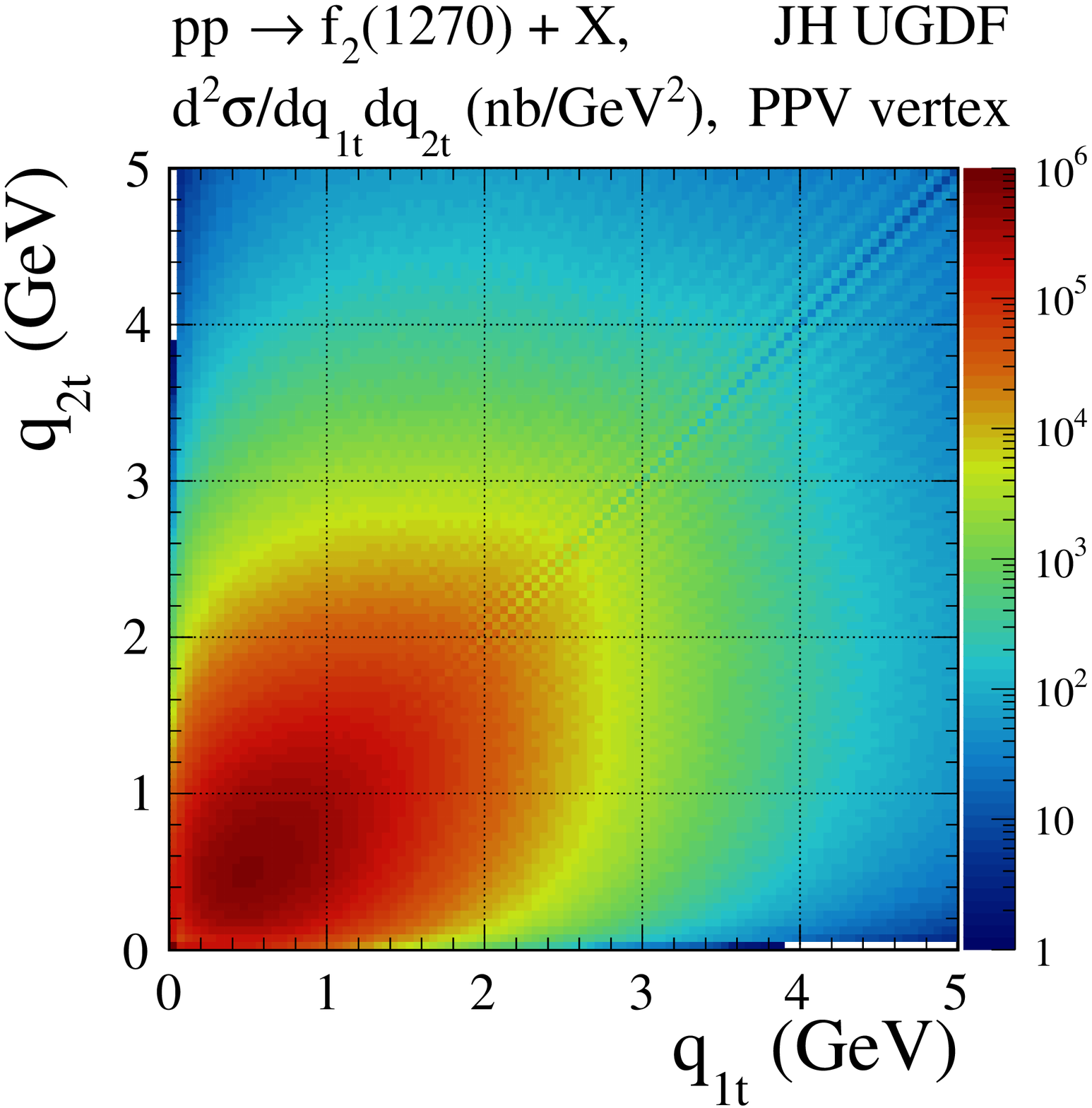}
\end{center}
\caption{\label{fig:dsig_dq1tdq2t}
\small
Two-dimensional distributions in gluon transverse momenta
for the JH UGDF and for two $g^{*}g^{*}f_{2}(1270)$ vertex prescription:
EMN (left panel) and PPV (right panel).
Here we used the dipole form factor (\ref{dipole}) 
with $\Lambda_D = m_{f_{2}}$.}
\end{figure}

In Fig.~\ref{fig:dsig_dq1tdq2t} we present
$d^{2}\sigma/dq_{1t} dq_{2t}$ for the EMN (left panel) 
and PPV (right panel) 
$g^{*}g^{*} \to f_{2}(1270)$ vertices.
Here the JH UGDF was used.
The maximal contributions
come from the region of rather small gluon transverse momenta
$q_{1t}, q_{2t} \lesssim 1$~GeV.
It is easy to check (numerically) that the larger-$p_t$ region 
($p_t > 2$~GeV) is sensitive
to $q_{1t}, q_{2t} > 1$~GeV where perturbative methods apply.
At low $p_t$ there is a nonnegligible contribution from the
nonperturbative region of UGDFs 
which is not under full theoretical control.
Here the gluon saturation effects may be potentially important.

We have checked that
\begin{equation}
\frac{d^2 \sigma_{\rm EMN}}{d q_{1t} d q_{2t}}
\left(\frac{d^2 \sigma_{ \rm PPV}}{d q_{1t} d q_{2t}} \right)^{-1} \to 1\,, \quad
{\rm for} \;\;q_{1t} \to 0 \;\;{\rm and} \;\;q_{2t} \to 0 \,,
\label{on_shell_limit}
\end{equation}
i.e. the two vertices are equivalent 
for both on-shell gluons.

\begin{figure}[!ht]
\begin{center}
\includegraphics[width=0.495\textwidth]{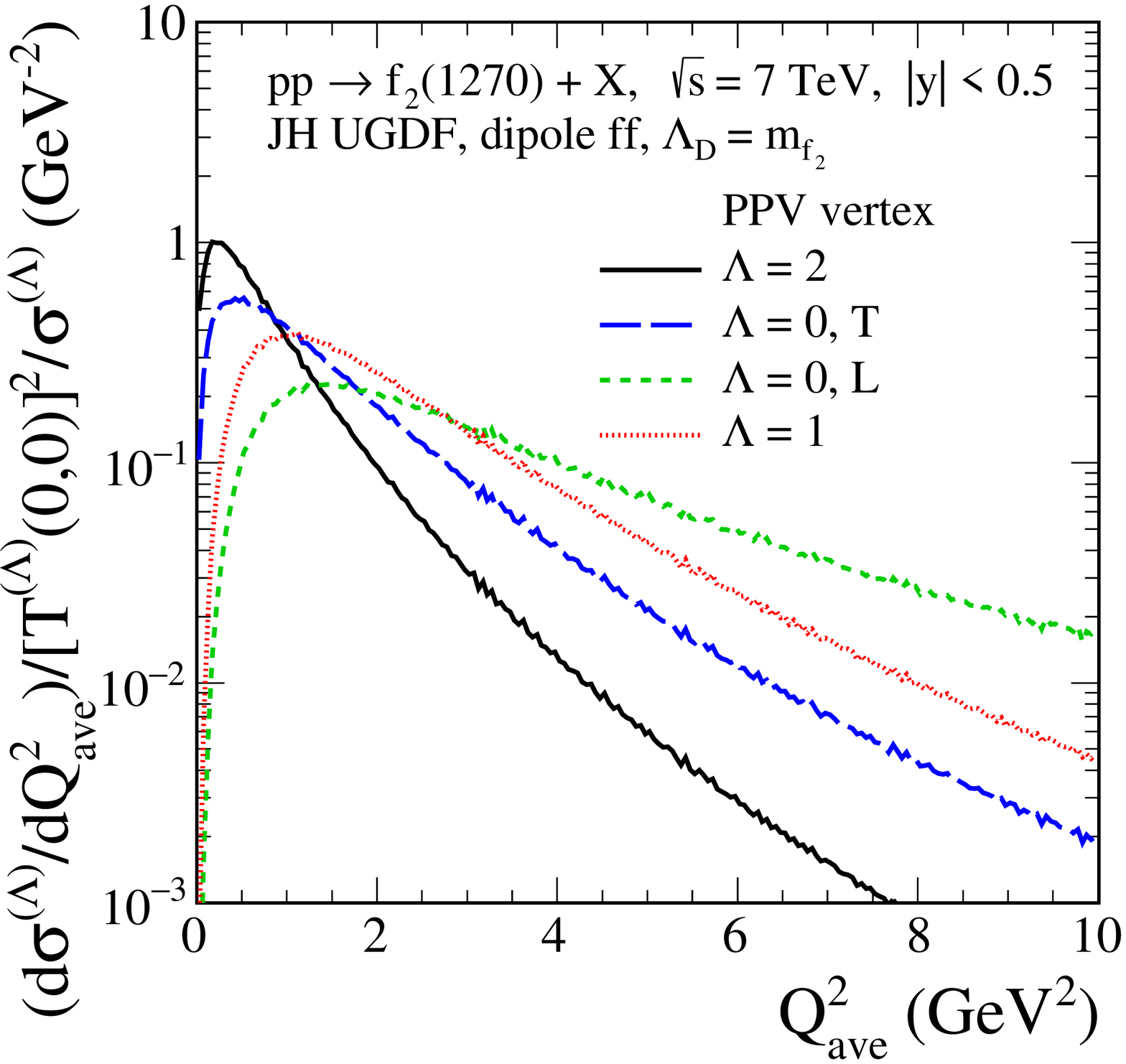}
\includegraphics[width=0.495\textwidth]{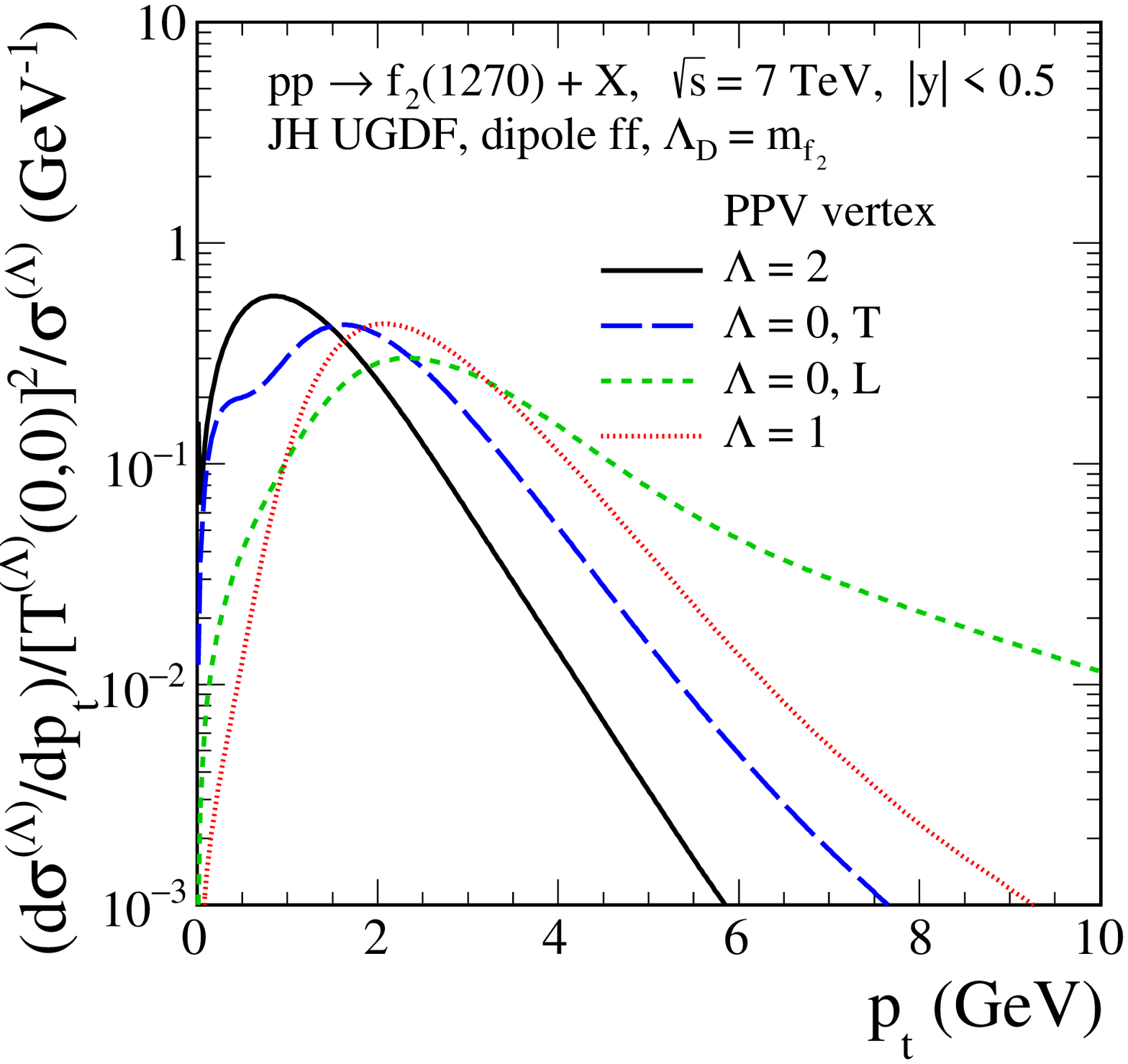}
\end{center}
\caption{\label{fig:dsig_dQ2ave}
\small
Distributions normalized as explained in the y-axis
in the averaged virtuality
\mbox{$Q^{2}_{\rm ave} = (Q_{1}^{2}+Q_{2}^{2})/2$} (left panel)
and in the $f_{2}(1270)$ meson transverse momentum (right panel).
Results for different 
$\Lambda = 0, 1, 2$ helicity terms in the $g^{*}g^{*} f_{2}$ vertex 
(\ref{PPV_vertex})--(\ref{TFF_aux}) using the same form of
vertex form factors $F^{(\Lambda)}(Q_{1}^{2}, Q_{2}^{2})$ (\ref{dipole}) with $\Lambda_D = m_{f_{2}}$ are shown.
In the calculation the JH UGDF was used.
}
\end{figure}

In Fig.~\ref{fig:dsig_dQ2ave} we show auxiliary
distributions to discuss a possible role of 
the remaining terms 
in the $g^{*}g^{*} f_{2}$ PPV vertex (\ref{PPV_vertex})
corresponding to helicities ($\Lambda = 0$,~L) and ($\Lambda = 1$).
Here we assumed the same $Q^{2}$ dependence of the form factor functions for all $\Lambda$ terms; 
see Eqs.~(\ref{TFF_aux}) and (\ref{dipole}).
The dominance of $\Lambda = 2$ term over $\Lambda = 0$
and $\Lambda = 1$ terms is certainly maintained at small values
of $Q^{2}_{\rm ave}$ and of $p_{t}$.
However, the situation changes drastically at large
gluon virtualities, i.e., 
the ($\Lambda = 1$) and ($\Lambda = 0$,~L) structures 
of the $g^{*}g^{*}f_{2}$ vertex 
become equally important for $p_t > 2$~GeV.

Note that from the analysis of the 
$\gamma^{*}(Q_{1}^{2})\gamma^{*}(Q_{2}^{2}) \to \pi\pi$ processes
performed in \cite{Hoferichter:2019nlq,Danilkin:2019opj}
it is clear that in the $f_{2}(1270)$ resonance region
the helicity-(0,~T) amplitude gives the dominant contribution
and the other helicity projections become 
increasingly important for larger virtualities.
From Fig.~5 of \cite{Hoferichter:2019nlq} and Fig.~3 of \cite{Danilkin:2019opj}
we can see that
for $Q_{1}^{2}$ fixed
the helicity-1 contribution increases with increasing $Q_{2}^{2}$ 
while helicity-(0,~L) contribution only slightly decreases.
The situation changes when both
photon virtualities are identical $Q_{1}^{2} = Q_{2}^{2}$ and large, 
i.e. then the helicity-(0,~L) component 
increases with increasing virtualities
and becomes even larger than the helicity-1 component.
This observation is consistent with our results 
presented in Fig.~\ref{fig:dsig_dQ2ave}.

The theoretical results
for the color-singlet gluon-gluon fusion contribution 
underestimate the ALICE data especially for low-$p_t$ region,
$p_t < 2$~GeV.
Does it mean that other mechanism(s) is (are) at the game? 

\begin{figure}[!ht]
\begin{center}
\includegraphics[width=0.495\textwidth]{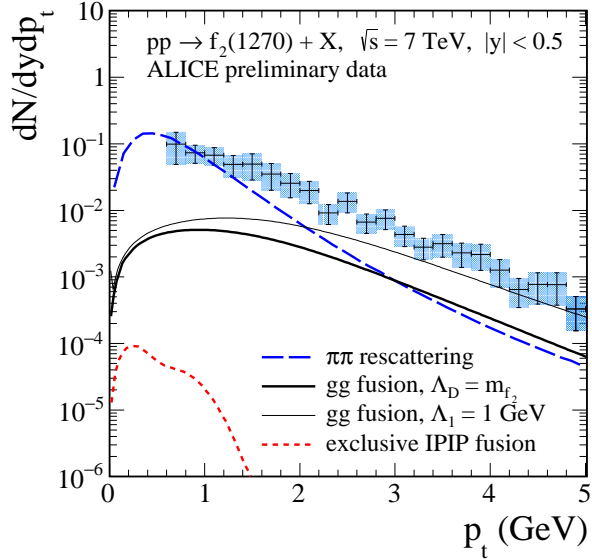}
\end{center}
\caption{\label{fig:low_pt}
\small
Results for the $\pi \pi$ rescattering mechanism (long-dashed line), for the $gg$-fusion mechanism (solid lines), 
and for the pomeron-pomeron fusion mechanism (dotted line)
together with the preliminary ALICE data from \cite{Lee:thesis}.
We show maximal contribution from the $\pi \pi$ rescattering 
as described in the main text.
The results for the $gg$-fusion contributions were calculated 
for the JH UGDF and for the PPV vertex 
[helicity-2 plus helicity-(0, T) terms]
and for two form factor functions (\ref{ffa}) 
(top solid line) and (\ref{dipole}) (bottom solid line).
The dotted line corresponds to the Born-level result 
for the $pp \to pp f_2(1270)$ process via pomeron-pomeron fusion.}
\end{figure}

In Fig.~\ref{fig:low_pt} we show the $\pi\pi$ rescattering contribution.
Clearly the $\pi \pi \to f_2(1270)$
rescattering effect cannot describe the region of $p_t > 2$~GeV,
where the $gg$-fusion mechanism is a possible explanation.
In addition, we present the Born result 
(without absorptive corrections important only 
when restricting to purely exclusive processes)
for the $pp \to pp f_2(1270)$ process
proceeding via the pomeron-pomeron fusion mechanism
calculated in the tensor-pomeron approach.
For details regarding this approach we refer to 
\cite{Ewerz:2013kda,Lebiedowicz:2013ika,Lebiedowicz:2016ioh,Lebiedowicz:2019por}.
In the calculation we take 
the pomeron-pomeron-$f_2(1270)$ coupling parameters from 
\cite{Lebiedowicz:2019por}.

\section{Conclusions}

In the present paper we have discussed production of $f_2(1270)$
tensor meson in proton-proton collisions. 
Two different approaches 
for the $\gamma^* \gamma^* \to f_{2}(1270)$ vertex,
according to EMN (\ref{EMN_vertex}) 
and PPV (\ref{PPV_vertex}) parametrizations,
have been considered. 
We have discussed their equivalence for both on-shell
photons. 
We have checked that the energy-momentum tensor vertex, 
proposed in \cite{FillionGourdeau:2007ee} 
[see Eq.~(A1) of \cite{FillionGourdeau:2007ee}], 
is equivalent to $\Gamma^{(2)}$
in the EMN vertex [see Eq.~(\ref{Gam2})]
when ignoring the coupling constants.
The coupling constants have been fixed by the Belle data
for $\gamma \gamma \to f_2(1270) \to \pi \pi$.
Then, the $g^* g^* \to f_2(1270)$ vertices have been obtained by replacing
electromagnetic coupling constant by the strong coupling constant,
modifying color factors and assuming a simple flavor structure of the $f_2(1270)$ isoscalar meson.

We have performed our calculation of the cross section for 
$p p \to f_2(1270) + X$ within the $k_t$-factorization approach.
Two different unintegrated gluon distributions 
from the literature have been used.
We have discussed corresponding uncertainties.


Our results have been compared to preliminary
ALICE data presented in \cite{Lee:thesis}.
We have taken into account only the case when both gluons are transverse.
At low $f_2(1270)$ transverse momenta
the helicity-2 ($\Lambda = 2$) contribution
dominates, while the helicity-0 ($\Lambda = 0$,~T) is small,
almost negligible, but competes
with the $\Lambda = 2$ and even dominates 
at larger transverse momenta of $f_2(1270)$. 
In the PPV formalism there could be also 
$\Lambda = 0$,~L and $\Lambda = 1$
contributions which are difficult to fix 
by available data.

It has been shown that the results strongly depend on the form of
the vertex form factor $F(Q_{1}^{2}, Q_{2}^{2})$. 
With the GVDM form factor used previously
in $\gamma^* \gamma \to f_2(1270)$ fusion 
\cite{Achasov:1985ad,Achasov:2015pha}
one cannot describe the preliminary ALICE data.
We have tried also other choices. With plausible form
factor [e.g., dipole ansatz (\ref{dipole}) 
with $\Lambda_{D} \simeq m_{f_{2}}$, 
factorized ansatz (\ref{ffa}) with $\Lambda_{1} \simeq 1$~GeV] 
one can describe the data for $p_t > 2$~GeV but it seems
impossible to describe the low-$p_t$ data.
Clearly some mechanism at low-$p_t$ must be in the game there.

We have shown that the final state $\pi \pi$ rescattering may be 
the missing candidate. A simple empirical model has been proposed.
Adjusting corresponding probability for the $\pi \pi \to f_2(1270)$
rescattering and the $\Lambda_D$ parameter in the dipole form factor 
for the $g^* g^* \to f_2$ vertex we have been able to describe 
the preliminary ALICE data.

The gluon saturation is expected at low $x_{1}$ and $x_{2}$
i.e. automatically rather low transverse momenta of $f_{2}(1270)$
where most probably the $\pi \pi$ rescattering dominates,
which does not allow observation of saturation.

We have calculated also 
the exclusive production of $f_{2}(1270)$ meson via
the pomeron-pomeron fusion mechanism
with the parameters found in our previous analysis
for the exclusive reaction $pp \to pp \pi^{+}\pi^{-}$.
This contribution is concentrated
at small $f_{2}(1270)$ transverse momenta 
but its role is rather marginal.

Our calculation suggest that the gluon-gluon 
fusion may be the dominant
mechanism of the $f_2(1270)$ production 
at larger transverse momenta, $p_t > 3$~GeV. 
Other mechanisms are of course not excluded 
but it is clear that the gluon-gluon fusion 
is a very important mechanism which cannot 
be ignored in the analysis.

\section*{Acknowledgments} 
We are indebted to Otto Nachtmann for a discussion of their approach.
We are indebted to Graham Richard Lee for the discussion of the experimental
data presented in his Ph.D. thesis \cite{Lee:thesis}. 
A discussion with Philip Ilten on production of the $f_2(1270)$ meson 
in the current version of {\tt PYTHIA} is acknowledged. 
We are also indebted to Marius Utheim
for explaining plans for including final state rescattering effects 
in the {\tt PYTHIA} code for production of $\pi \pi$ resonances.
This study was partially supported 
by the Polish National Science Centre under grant No.
2018/31/B/ST2/03537 and by the Center for Innovation and
Transfer of Natural Sciences and Engineering Knowledge in Rzesz\'ow (Poland).


\end{document}